\documentclass[prl,twocolumn,superscriptaddress]{revtex4}
\usepackage{amsmath,graphicx}
\usepackage{epsfig}
\usepackage{paralist}
\usepackage{amsmath,amssymb,mathrsfs}
\usepackage{underscore}
\usepackage{natbib}
\usepackage{color}
\usepackage{dsfont}
\usepackage{bm}
\usepackage[colorlinks,linkcolor=blue,anchorcolor=blue,urlcolor=blue,citecolor=blue]{hyperref}

\begin{document}

\title{Superradiant phase transition with cavity assisted dynamical spin-orbit coupling}

\author{Ying Lei}
\affiliation{School of Physics, Huazhong University of Science and Technology, Wuhan 430074, China}

\author{Shaoliang Zhang}
\email{shaoliang@hust.edu.cn}
\affiliation{School of Physics, Huazhong University of Science and Technology, Wuhan 430074, China}
\affiliation{State Key Laboratory of Quantum Optics and Quantum Optics Devices, Shanxi University, Taiyuan, 030006,China}

\date{\today}

\begin{abstract}
Superradiant phase transition represents an important quantum phenomenon that shows the collective excitations based on the coupling between atoms and cavity modes. The spin-orbit coupling is another quantum effect which induced from the interaction of the atom internal degrees of freedom and momentum of center-of-mass. In this work, we consider the cavity assisted dynamical spin-orbit coupling which comes from the combination of these two effects. It can induce a series of interesting quantum phenomena, such as the flat spectrum and the singularity of the excitation energy spectrum around the critical point of quantum phase transition. We further discuss the influence of atom decay and nonlinear coupling to the phase diagram. The atom decay suppresses the singularity of the phase diragram and the nonlinear coupling can break the symmetric properties of the phase transition. Our work provide the theoretical methods to research the rich quantum phenomena in this dynamic many-body systems. 
\end{abstract}

\maketitle

\section{introduction}

The cavity quantum electrodynamics (CQED) has a wide range of applications in quantum information and quantum simulations, because the interaction between hyperfine spin states of atom and quantum photon field has a great enhancement by the cavity. A famous theoretical model which used to describe this interaction is called Dicke model \cite{Dicke}. It is well known to undergo a quantum phase transition from normal phase to superradiant phase called "superradiant phase transition" \cite{Lieb1, YKWang}. Over the last few years, based on the development of experimentally techniques, there are a series of works focus on researching the properties of this system \cite{FDimer1, DNagy, Viehmann, Bastidas,bhaseen} and superradiant phase transition has also be observed experimentally first by the group at ETH \cite{Baumann1,Mottl, Brennecke} then many other groups \cite{Schimdt, Klinder, zqzhang1}. Many interesting quantum phenomena has been observed such as supersolid-like phase, and research about their applications has also been developed rapidly \cite{gietka}.

The standard Dicke model only includes the coupling between cavity modes and atomic internal degrees of freedom. But in ultracold atomic system, the coupling between the internal degrees of freedom and the atomic center-of-mass(COM) motion, which is called the spin-orbit coupling (SOC) effect, has attracted a lot of interests in past two decades\cite{SOC1, SOC2, SOC3}. With the SOC effect, a lot of interesting quantum phenomena, especially the phenomena related to the topological physics, such as the synthetic gauge fields \cite{OFL1, OFL2}, chiral edge currents \cite{edgecurrent1, edgecurrent2}, has been observed experimentally. In ultracold atom systems, the SOC effect can be realized by coupling two different hyperfine spin states using two-photon Raman process which also accompanied by the transfer of COM momentum. In systems with static SOC, the translation invariance is conserved and the energy dispersion has some anomalous properties because of the SOC effect. It is the source of many interesting quantum phenomena such as the stripe phase \cite{cjwang} and algebra quantum liquids \cite{qizhou1}, etc. Recently, the cavity-assisted dynamical spin-orbit coupling which induced by a superradiant phase transition has been realized in experiment by replacing one of the Raman laser with an optical cavity \cite{Baden,zqzhang2,BLLev1}. In this system, the cavity modes not only couple with the hyperfine spin states of atoms, but also interact with the atomic COM momentum. The interplay of cavity modes, atomic internal and external degrees of freedom are expected to lead to novel quantum phenomena \cite{luzhou} and pave a way for the research of quantum simulation in topological matters and many-body physics.

Our propose in this work is to understand the interplay of these degrees of freedom.  In the system with cavity-assisted dynamical SOC, the SOC comes from the collective excitation of cavity photons and all atoms --- {\it the superradiance}, which is very different from the coupling between Raman laser and single atom in static SOC. Although the translation invariance is still conserved, the energy dispersion is very different and some exotic quantum phenomena emerges, such as the dynamical SOC induced flat spectrum. In this system, the superradiant phase transition has some interesting properties because of the interplay among cavity modes, hyperfine spin states and the COM momentum of atoms, which is very different from the standard Dicke model. There is a first order phase transition and the energy spectrum of polaritons has singularity at the critical point. In this work, we also discuss the influence of atom decay and nonlinear coupling to the phase diagram. We find that the atom decay suppresses the singularity of the phase diagram which result in the first order phase transition is hard to be observed experimentally, and the nonlinear coupling can break the symmetric form of the superradiant phase transition.

Our paper is organized as follows. In Sec. II we'll give a brief introduction about the cavity-assisted dynamical SOC model. Then we calculate the energy spectrum and find the flat spectrum at critical point of superradiant phase transition. We also draw the phase diagram and discuss the first order phase transition. In Sec. III we discuss the energy spectrum of polaritons and find the singularity induced by the flat spectrum. In Sec. IV we discuss the influence of atom decay and cavity photon loss to the phase diagram. We find that in this dissipative system, the singularity of the phase diagram will disappear because of the atom decay. In Sec. V we discuss the influence of nonlinear coupling to the phase diagram. The Sec. VI is devoted to our conclusions.

\section{Cavity-assisted spin-orbit coupling}

\begin{figure}[tph]
\begin{center}
{\includegraphics[width=0.4 \textwidth]{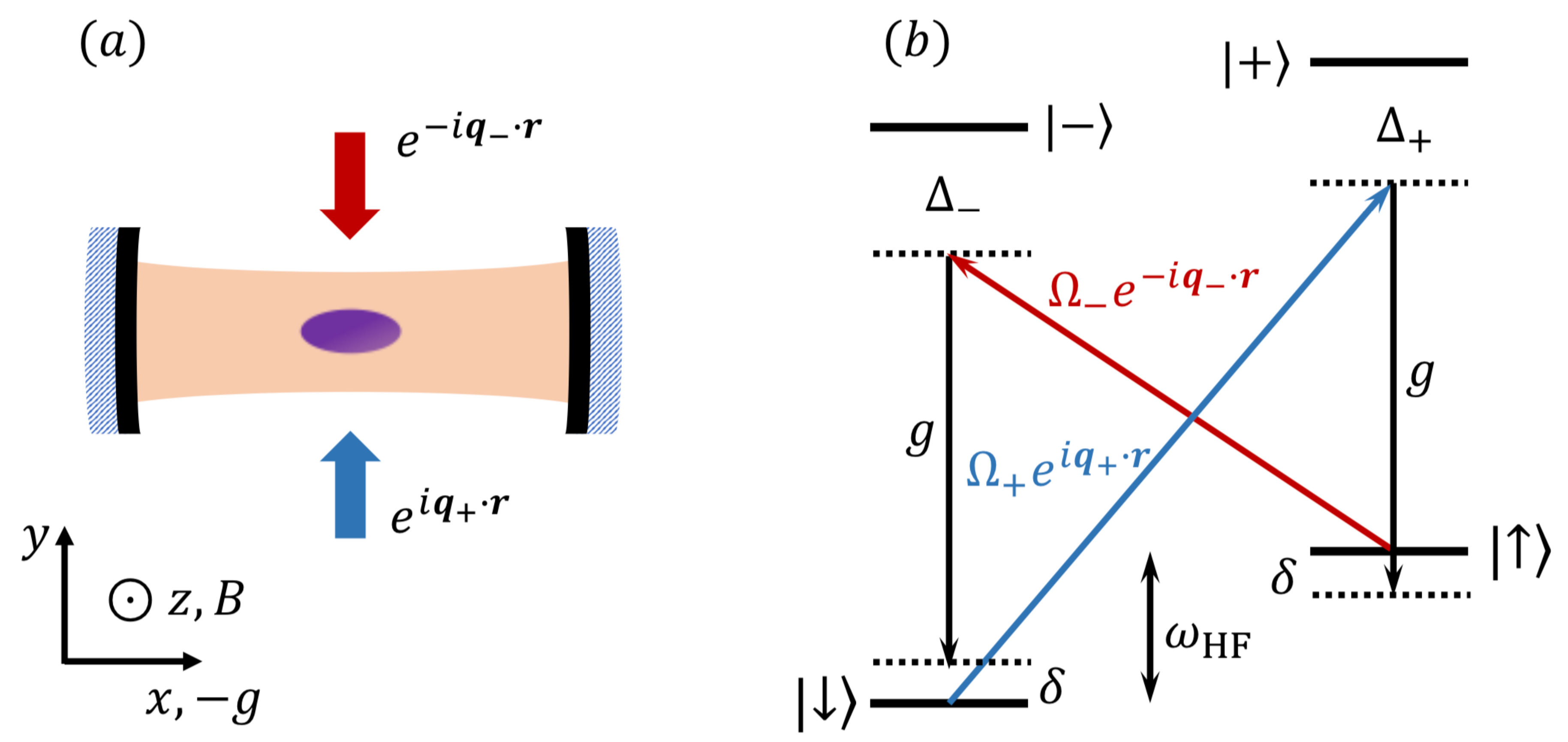}}
\caption{The schematic diagram of our proposal. (a) Experimental setup; (b) Level diagram of atom and light configuration.}
\end{center}
\end{figure}

We consider the atoms with two hyperfine spin states (denoted as $|\uparrow\rangle$ and $|\downarrow\rangle$) confined in an optical cavity, as shown schematically in Fig. 1. The cavity supports a single-mode quantum field with frequency $\omega$, two counter-propogating external lights (red and blue arrows) with frequency $q_+=q_-=q_r$ produces additional classical lights with different frequencies, which couple different hyperfine spin states respectively. Each of these two classical lights and cavity quantum field induce a two-photon Raman transition between different hyperfine spin states, and spontaneously transfer COM momentum of $\pm q_r$ along the direction perpendicular to the cavity axis in average. In the experiment of R. M. Kroeze et. al\cite{BLLev1}, the discretized momentum distribution and the spatial modulation along cavity axis have been observed. But in this work, we just focus on the properties perpendicular to the cavity axis and the spatial modulation can be ignored along this direction. After an unitary transformation, the effective Hamiltonian can be written as \cite{FDimer1, BLLev1}

\begin{equation}
\begin{split}
\mathcal{H}_\mathrm{eff}=&\sum^N_{j=1}\Big(\frac{{\bf p}^2_j}{2m}+\frac{{\bf p}_j\cdot{\bf q}_r}{m}\sigma^j_z\Big)+\delta S_z-\frac{U}{N}a^\dag aS_z \\
&+\frac{\lambda}{\sqrt{N}}(a^\dag+a)(S_++S_-)+\omega a^\dag a
\label{socham}
\end{split}
\end{equation}
where $\omega$, $\delta$, $\lambda$ and $U$ are the effective cavity frequency, energy detuning between hyperfine spin states, coupling amplitude and dispersive cavity shift. In equation (\ref{socham}), ${\bf p}_j$ is the momentum operator of the $j$-th atoms, $S_\alpha=\sum^N_{j=1}\sigma^j_\alpha$ where $\alpha=x,y,z$ and $S_\pm=S_x\pm iS_y$.
In most part of our work, we'll set the nonlinear atom-photon interaction $U=0$ to simplify our discussion, the influence of nonzero $U$ will be discussion in the Sec. V. 

In this work, we only consider the case all atoms are condensed with the same COM momentum ${\bf p}$. So ${\bf p}_j$ in equation (\ref{socham}) can be simplified as a number instead of an operator.  To further simplify our discussion, we can set the spin-orbit coupling along $y$ direction and ignore the influence of other direction, the above Hamiltonian can further be simplified as
\begin{equation}
\begin{split}
\mathcal{H}_\mathrm{eff}=&\frac{p^2}{2m}N+\Big(\frac{pq_r}{m}+\delta\Big)S_z+\omega a^\dag a \\
&+\frac{\lambda}{\sqrt{N}}(a^\dag+a)(S_++S_-)
\label{effhamsocmanybody}
\end{split}
\end{equation} 

Different with the original Dicke model, the Hamiltonian (\ref{effhamsocmanybody}) is also dependent on the COM momentum of atoms. The SOC effect doesn't break the translation invariant symmetry, and the commutation relation $[p,\mathcal{H}_\mathrm{eff}]=0$ is kept, which means the COM momentum of atom $p$ is still a good quantum number, and the eigenstates of the atoms are still plane waves. Because of the inversion symmetry, the Hamiltonian (\ref{effhamsocmanybody}) is conserved with translation of $p\rightarrow -p$, $\delta\rightarrow -\delta$ and $S_z\rightarrow -S_z$, which means the ground state has the same energy for different sign of $\delta$, so only $\delta\ge 0$ need to be considered. Define the effective energy detuning between $|\uparrow\rangle$ and $|\downarrow\rangle$ states as $\omega_0\equiv\frac{pq_r}{m}+\delta$ and ignores the influence of the kinetic energy term, the Eq.(\ref{effhamsocmanybody}) becomes to the standard Dicke model as
\begin{equation}
\mathcal{H}_\mathrm{Dicke}=\omega_0S_z+\omega a^\dag a+\frac{\lambda}{\sqrt{N}}(a^\dag+a)(S_++S_-)
\label{dickemodel}
\end{equation}
With the analytical calculation in the work of C. Emary and T. Brandes\cite{EmaryB}, one can get the eigenenergy and corresponding wave functions for different phase, the critical point of superradiant phase transition is at $\lambda_\mathrm{cr}=\sqrt{\omega\omega_0}/2$ in thermodynamic limit.

Consider the Hamiltonian (\ref{effhamsocmanybody}), for an arbitrary coupling strength $\lambda$ and energy detuning $\delta$, one can get the energy spectrum with different COM momentum $p$ first. Because of the relative amplitude of $\lambda$ and $\omega_0$ at different $p$, the corresponding eigenstates should be superradiant states or normal states due to different COM momentum $p$. (i) For $-\frac{m}{q_r}\big(\frac{4\lambda^2}{\omega}+\delta\big)< p<\frac{m}{q_r}\big(\frac{4\lambda^2}{\omega}-\delta\big)$, the relation $\lambda>\lambda_\mathrm{cr}$ can be satisfied, so the eigenstate is superradiant state and the corresponding energy can be expressed as
 \begin{equation}
 \frac{E_G}{N}=\frac{8m\lambda^2-\omega q^2_r}{16m^2\lambda^2}p^2-\frac{\omega\delta q_r}{8m\lambda^2}p-\frac{\lambda^2}{\omega}-\frac{\omega\delta^2}{16\lambda^2}
 \label{genergyssr}
 \end{equation} 
 (ii) For the other $p$, we have $\lambda<\lambda_\mathrm{cr}$, the corresponding eigenstate is normal state and the energy should be
 \begin{equation}
\frac{E_G}{N}=\frac{p^2}{2m}\mp\frac{1}{2}\Big(\frac{pq_r}{m}+\delta\Big)
\label{genergynormal}
\end{equation}
where $\mp$ correspond to $p\ge \frac{m}{q_r}\big(\frac{4\lambda^2}{\omega}-\delta\big)$ and $p\le -\frac{m}{q_r}\big(\frac{4\lambda^2}{\omega}+\delta\big)$ respectively. 

\begin{figure}[tph]
\begin{center}
{\includegraphics[width=0.48 \textwidth]{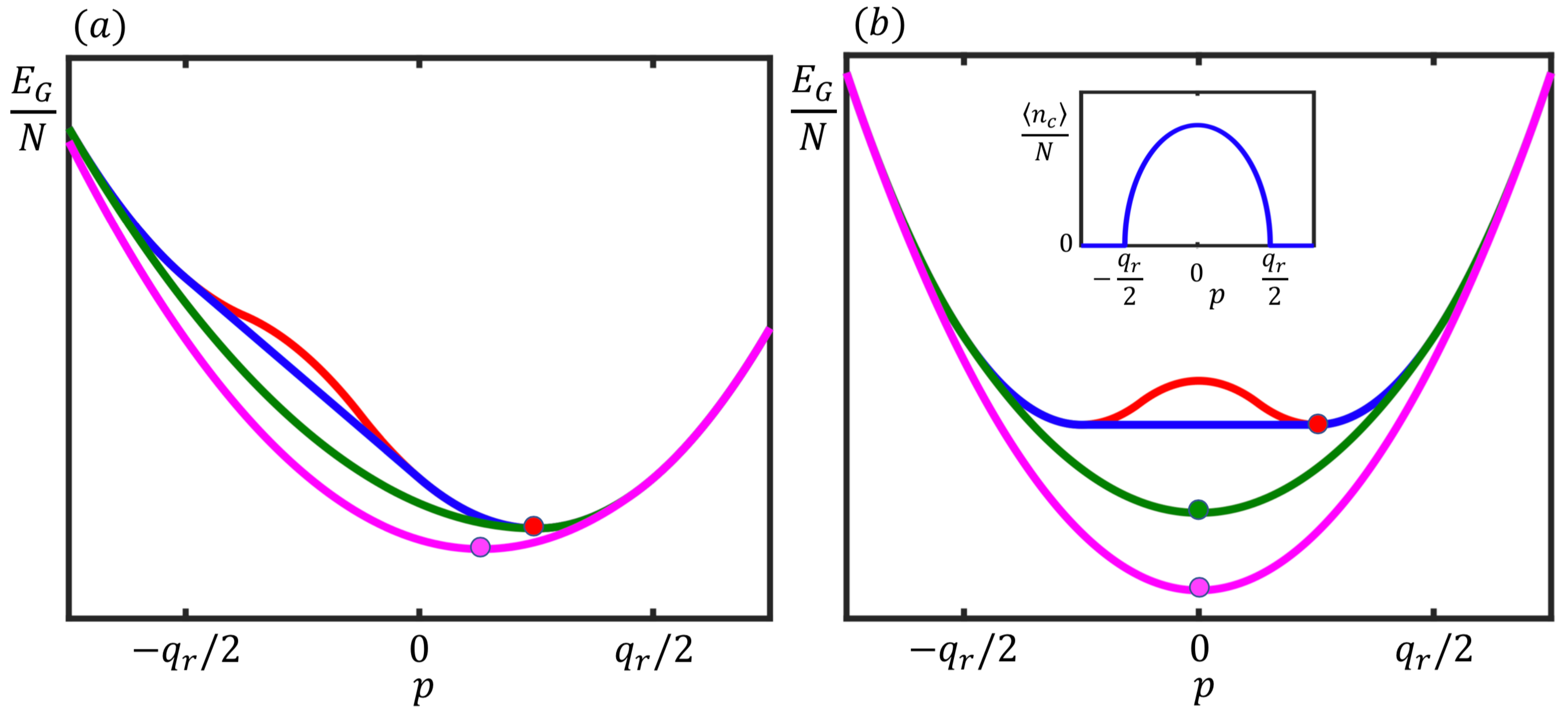}}
\caption{The energy spectrum for different COM momentum. Different color lines correspond to different $\lambda$. (a) $\delta>0$, $\lambda^2<\omega E_r/4$ for the red line, $\lambda^2=\omega E_r/4$ for the blue line, $\lambda^2=\frac{\omega}{4}\big(E_r+\delta\big)$ for the green line and $\lambda^2>\frac{\omega}{4}\big(E_r+\delta\big)$ for the magenta line. The red dot is the energy minimum for the first three cases, which is at $p_\mathrm{min}=q_r/2$, the magenta dot is the energy minimum for the last case, which is at $p_\mathrm{min}<q_r/2$; (b) $\delta=0$.$\lambda^2<\omega E_r/4$ for the red line, $\lambda^2=\omega E_r/4$ for the blue line, $\lambda^2>\omega E_r/4$ for the green and magenta lines. The red dot is the energy minimum for the first two cases, which is at $p_\mathrm{min}=q_r/2$, the green and magenta dots are the energy minimum for the last two cases, which is at $p_\mathrm{min}=0$. The inset shows the expectation value of cavity photon number $\langle n_c\rangle/N$ for the eigenstates with different $p$ where $\delta=0$ and $\lambda=\sqrt{\omega E_r}/2$. }
\end{center}
\end{figure}

The energy spectrum of the eigenstates with different COM momentum $p$ is shown in Fig. 2. The ground state is at the minimum of these energy spectrum and is pointed out with fixed dot. We find that with the increase of the coupling strength $\lambda$, the energy of superradiant state is decrease  until two state energy are the same, there is a superradiant phase transition in this system at $\lambda^2_c=\omega\big(E_r+\delta\big)/4$. The phase transition need to be discussed in two different cases. (i). $\delta>0$. In this case, for $\lambda^2\le\lambda^2_c=\omega\big(E_r+\delta\big)/4$ (where $E_r=q^2_r/(2m)$ is the recoil energy), the ground state is normal state at momentum $p_\mathrm{min}=q_r/2$. For $\lambda^2>\omega\big(E_r+ \delta \big)/4$, the ground state is superradiant state and the corresponding momentum is $p_\mathrm{min}=\frac{\omega\delta q_r}{2(4\lambda^2-\omega E_r)}$, which moves asymptotically to the point $p=0$ with the increase of $\lambda$. (ii). $\delta=0$. In this extreme case, the ground state is also normal state when $\lambda^2<\omega E_r/4$ with momentum $p_\mathrm{min}=q_r/2$. If $\lambda^2>\omega E_r/4$, the ground state is superradiant state with momentum $p_\mathrm{min}=0$. At critical point where $\lambda^2=\omega E_r/4$, the energy spectrum is flat in the range of $q\in[-q_r/2, q_r/2]$, as shown the blue line in Fig. 2(b), which is one of the most interesting property in this system. This flat spectrum comes from the interplay of the cavity modes, atomic internal and external degrees of freedom. 
The coupling energy between cavity photons and hyperfine spin states and the kinetic energy of atoms which both having the quadratic form of COM momentum $p$ cancel each other out at the critical point. 

\begin{figure}[tph]
\begin{center}
{\includegraphics[width=0.5 \textwidth]{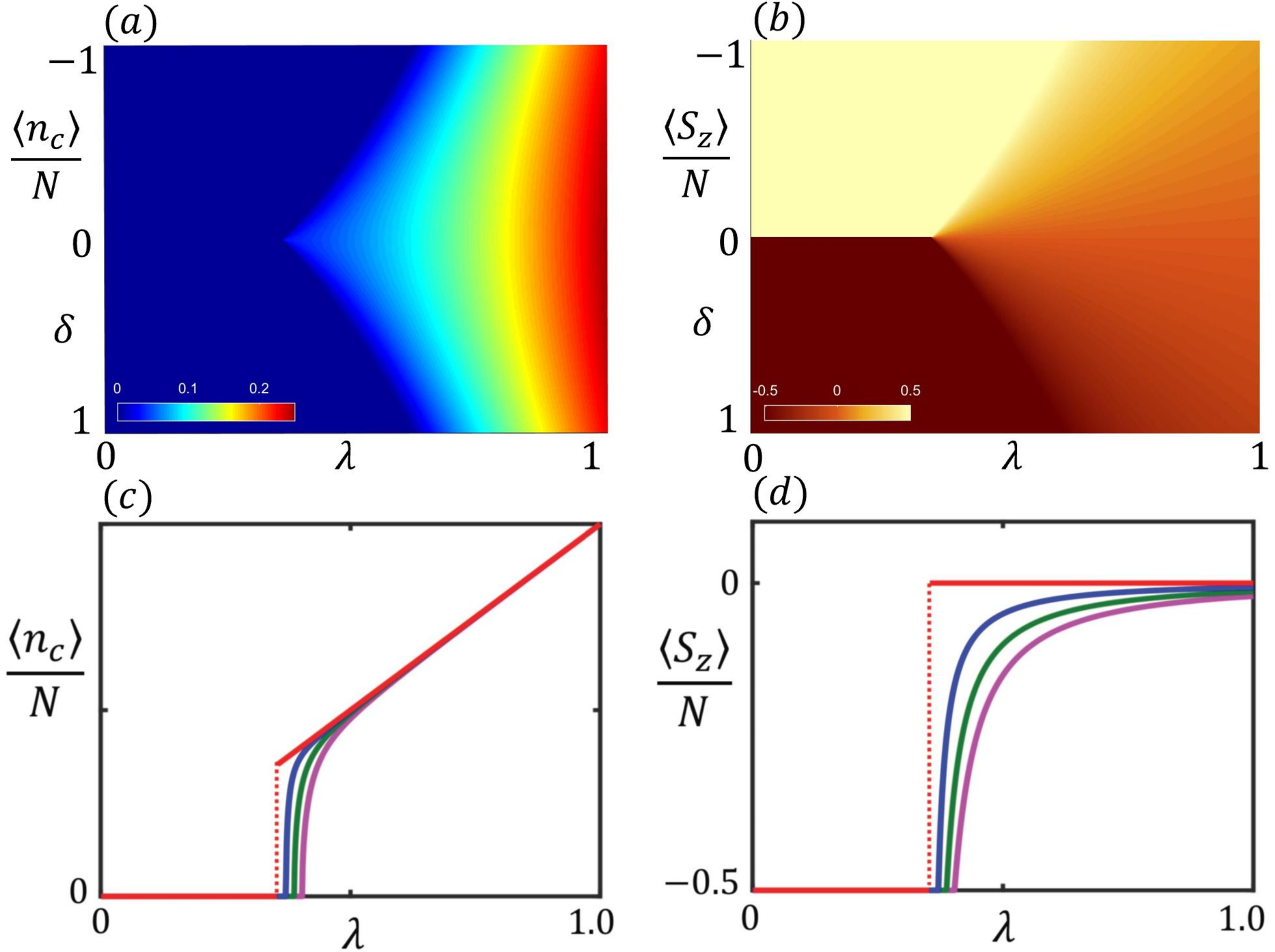}}
\caption{Phase diagram. (a). The expectation value of cavity photon number $\langle n_c\rangle$. (b). The expectation value of $\langle S_z\rangle$. (c). The expectation value of cavity photon number for different $\delta$. The red, blue, green, magenta line correspond to $\delta=0$, $\delta=0.1$, $\delta=0.2$ and $\delta=0.3$ respectively. (c). The expectation value of $\langle S_z\rangle$ for different $\delta$. The red, blue, green, magenta line correspond to $\delta=0$, $\delta=0.05$, $\delta=0.1$ and $\delta=0.15$ respectively. }
\end{center}
\end{figure}

With the above discussion, one can also draw the phase diagram in thermodynamic limit, as shown in Fig. 3. Fig. 3(a) and (b) show the expectation of cavity photons number $\langle n_c\rangle=\langle a^\dag a\rangle$ and magnetic quantum number $\langle S_z\rangle$ with different $\delta$ and $\lambda$ respectively. From the phase diagram, one can easily find there is a singularity where both expectation values have discontinuity along $\delta=0$, as shown the red lines in Fig. 3(c) and Fig. 3(d), which means the system has the first order phase transition. All these novel phenomena come from the flat spectrum at critical point. The flat spectrum means the system is highly degenerate at this critical point. But all these eigenstates have different number of cavity photons and magnetic quantum number because the effective detuning $\omega_0$ are different in these states with different $p$, as shown in inset of Fig. 2(b). The expectation of cavity photon number is the minimum $\langle n_c\rangle=0$ at $p=\pm q_r/2$ but the maximum $\langle n_c\rangle/N=E_r/(4\omega)$ at $p=0$. After passing the critical point, the momentum of COM of ground state is suddenly changed from $p_\mathrm{min}=\pm q_r/2$ to $p_\mathrm{min}=0$, the expectation of cavity photon number increases from the finite value $\langle n_c\rangle/N=E_r/(4\omega)$ instead of zero. We should also point out that when $\delta=0$, the ground state energy has two fold degeneracy at $p=\pm q_r/2$ in normal phase because of the symmetry, which means that all spins occupying $|\uparrow\rangle$ and $|\downarrow\rangle$ will have the same energy. In the red line in Fig. 3(d), we plot the value $\langle S_z\rangle/N$ as $-1/2$ instead of $1/2$ when $\lambda^2<\omega E_r/4$, it can be considered as a spontaneous symmetry breaking.

\section{The properties of excitations}

To explore the properties of the superradiant phase transition further, we estimate the energy of the ground state. When $\delta\ne0$, the ground state energy can be written as
\begin{equation}
\epsilon(\delta,\lambda)=\left\{\begin{array}{c} -\frac{E_r}{4}-\frac{\delta}{2}\,\,\,(\lambda\le\lambda_c) \\ -\frac{\omega^2\delta^2E_r}{16\lambda^2(4\lambda^2-\omega E_r)}-\frac{\lambda^2}{\omega}-\frac{\omega\delta^2}{16\lambda^2}\,\,\,(\lambda>\lambda_c)  \end{array}\right.
\end{equation} 
where $\lambda^2_c=\frac{\omega}{4}\Big(E_r+\delta\Big)$. With a direct calculation, one can easily estimate the second order derivation of the energy with $\delta$ and $\lambda$ and find that there is a discontinuity around the phase boundary, so the system has a second order phase transition when $\delta\ne 0$, as the standard Dicke model.

In superradiant phase, the collective excitations of the atoms and cavity photons are called polaritons, which can be effectively equivalent to two independent harmonic oscillator modes \cite{EmaryB}. In standard Dicke model, one of oscillator modes approaches to zero from both sides with the asymptotic form $\propto |\lambda-\lambda_\mathrm{cr}|^{1/2}$, where $\lambda_\mathrm{cr}=\sqrt{\omega\omega_0}/2$ is the critical point of standard Dicke model. In the system with Hamiltonian (\ref{effhamsocmanybody}), for an arbitrary $\delta$, the excitation energy can be expressed as the function of $\lambda$ as\cite{TLWang}
\begin{equation}
2\epsilon^2_\pm=\omega^2+\frac{\omega^2_0}{\mu^2}\pm\sqrt{\Big(\omega^2-\frac{\omega^2_0}{\mu^2}\Big)^2+(4\lambda)^2\omega\omega_0\mu}
\label{polaritonen}
\end{equation}
where $\omega_0=\frac{p_\mathrm{min}q_r}{m}+\delta$ is the effective energy detuning between two hyperfine spin states, $\mu=1$ in normal phase and $\mu=\lambda^2_{c}/\lambda^2<1$ in superradiant phase. The energy of polaritons has the same asymptotic form with the standard Dicke model at the normal phase side when $\lambda\le\lambda_{c}(\delta)$. But at the superradiant phase side, the $\omega_0$ is the function of $\delta$ instead of a constant in standard Dicke model, the asymptotic form will change with $\delta$ as
\begin{equation}
\epsilon_-\sim\sqrt{\frac{64\lambda^3_0\Big(1+\frac{E_r}{\delta}\Big)}{\frac{16\lambda^4_0}{\omega^2}+\omega^2}}|\lambda-\lambda_{c}|^\frac{1}{2}
\end{equation}

\begin{figure}[tph]
\begin{center}
{\includegraphics[width=0.47 \textwidth]{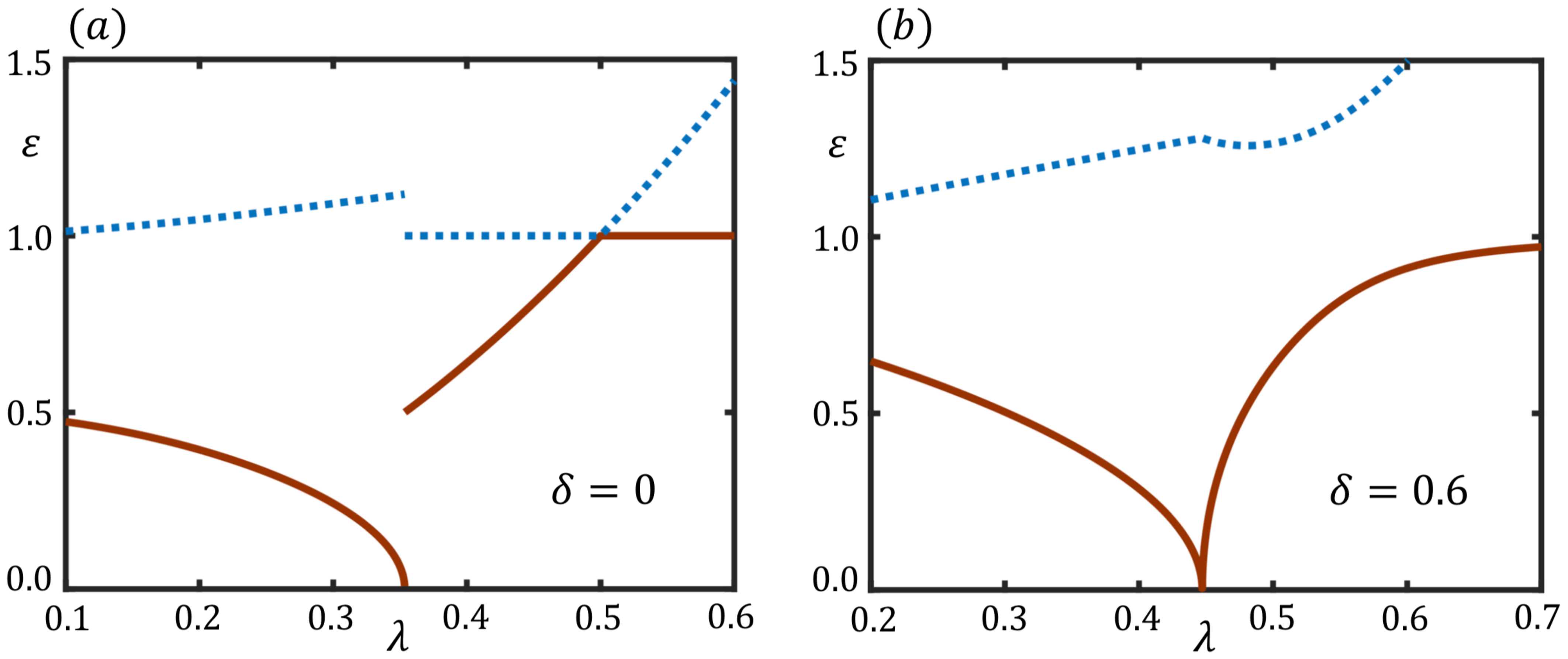}}
\caption{The excitation energies as the function of $\lambda$. The parameters are all scaled to $\omega=m=q_r=1$. The brown solid line and the blue dotted line correspond to $\epsilon_-$ and $\epsilon_+$ respectively. (a) $\delta=0$, the excitation energy is discontinuous. (b). $\delta=0.6$, the excitation energies is similar with standard Dicke model. }
\end{center}                                           
\end{figure}

With decrease of $\delta$, the energy spectrum becomes more and more steep. When $\delta=0$, the energy of polaritons (\ref{polaritonen}) has a singularity at the critical point as shown in Fig.4(a). This is the result of first order phase transition. We can also estimate the excitation energy analytically on the side of superradiant phase in this extreme case as $\epsilon_+=4\lambda^2/\omega$ and $\epsilon_-=\omega$, which means the atomic branch and photon branch are decoupled. The reason of this decoupling is the COM momentum in the superradiant phase side is $p=0$ when $\delta=0$, then the effective energy detuning of different hyperfine spin states is zero.

\section{The influence of atom decay}

In standard Dicke model (\ref{dickemodel}), the critical point of superradiant phase transition is severely dependent on the dissipation of cavity photons and atoms \cite{Gelhausen}. By considering the atomic decay and cavity loss, the master equation of the density matrix can be written as
\begin{equation}
\begin{aligned}
\dot{\rho}=-i[\mathcal{H}_\mathrm{eff}, \rho]&+\gamma \sum_{\ell=1}^{N}\left(\sigma^{\ell}_{-} \rho \sigma^{\ell}_{+}-\frac{1}{2}\left\{\sigma^{\ell}_{+} \sigma^{\ell}_{-}, \rho\right\}\right)\\&+\kappa\left[2 a \rho a^{\dagger}-\left\{a^{\dagger} a, \rho\right\}\right]
\label{mastereq}
\end{aligned}
\end{equation} 
The Lindblad operators include two parts, the first part describes the atom decay with rate $\gamma$ and the second part describes the cavity photon loss with decay rate $\kappa$. We should point out that the atom decay rate in above equation (\ref{mastereq}) is not the rate of spontaneous emission from excited state to ground state, because both of the hyperfine spin states are the ground states. This decay rate comes from the difference of spontaneous emission rate to these two ground states from excited states which have been eliminated adiabatically, such as the $|+\rangle$ and $|-\rangle$ states in Fig. 1(b). This type of decay will not induce the atom number loss but only the spin decoherence. If we ignore the momentum transfer in the spontaneous emission process in average, the COM momentum $p$ can still be a number. Then with the standard method in quantum optics, the Heisenberg-Langevin equations for the atomic and cavity photon operators can be written down and the steady state solutions can be estimated as shown in \cite{Gelhausen}. Then with the similar method in the Section II, one can also estimate the steady state with $p$ which has the lowest energy expectation and draw the phase diagram. The critical point of the superradiant phase transition has be changed into:
\begin{equation}
\lambda_c=\frac{1}{4}\sqrt{\frac{(\gamma^2+4\omega^2_0)(\kappa^2+\omega^2)}{\omega_0\omega}}
\end{equation}
where $\omega_0=\frac{q^2_r}{2m}+|\delta|$.The expectation value of operators $\langle a\rangle$, $\langle\sigma_+\rangle$ and $\langle \sigma_z\rangle$ in superradiant phase can be expressed as
\begin{equation}
\begin{aligned}
\langle a\rangle &=\pm \frac{\sqrt{\frac{\kappa^{2}+\omega^{2}}{\omega}} \sqrt{\omega_0\left(1-\frac{(\gamma^2+4\omega_0^2)(\kappa^2+\omega^2)}{16\lambda^2\omega\omega_0}\right)}}{\sqrt{2}\left(-\omega+i \kappa\right)} \\
\left\langle\sigma_{+}\right\rangle &=\pm \frac{\sqrt{\omega_0\left(\frac{\lambda^2\omega}{\kappa^2+\omega^2}-\frac{\gamma^2+4\omega_0^2}{16\omega_0}\right)}(\kappa^2+\omega^2)}{\sqrt{2} \lambda^2\omega}\left(\frac{1}{2}-\frac{i\gamma}{4\omega_0}\right) \\
\left\langle\sigma^{z}\right\rangle &=-\frac{(\gamma^2+4\omega_0^2)(\kappa^2+\omega^2)}{16\lambda^2\omega\omega_0}
\label{meanvalue}
\end{aligned}
\end{equation} 

\begin{figure}[tph]
\begin{center}
{\includegraphics[width=0.47 \textwidth]{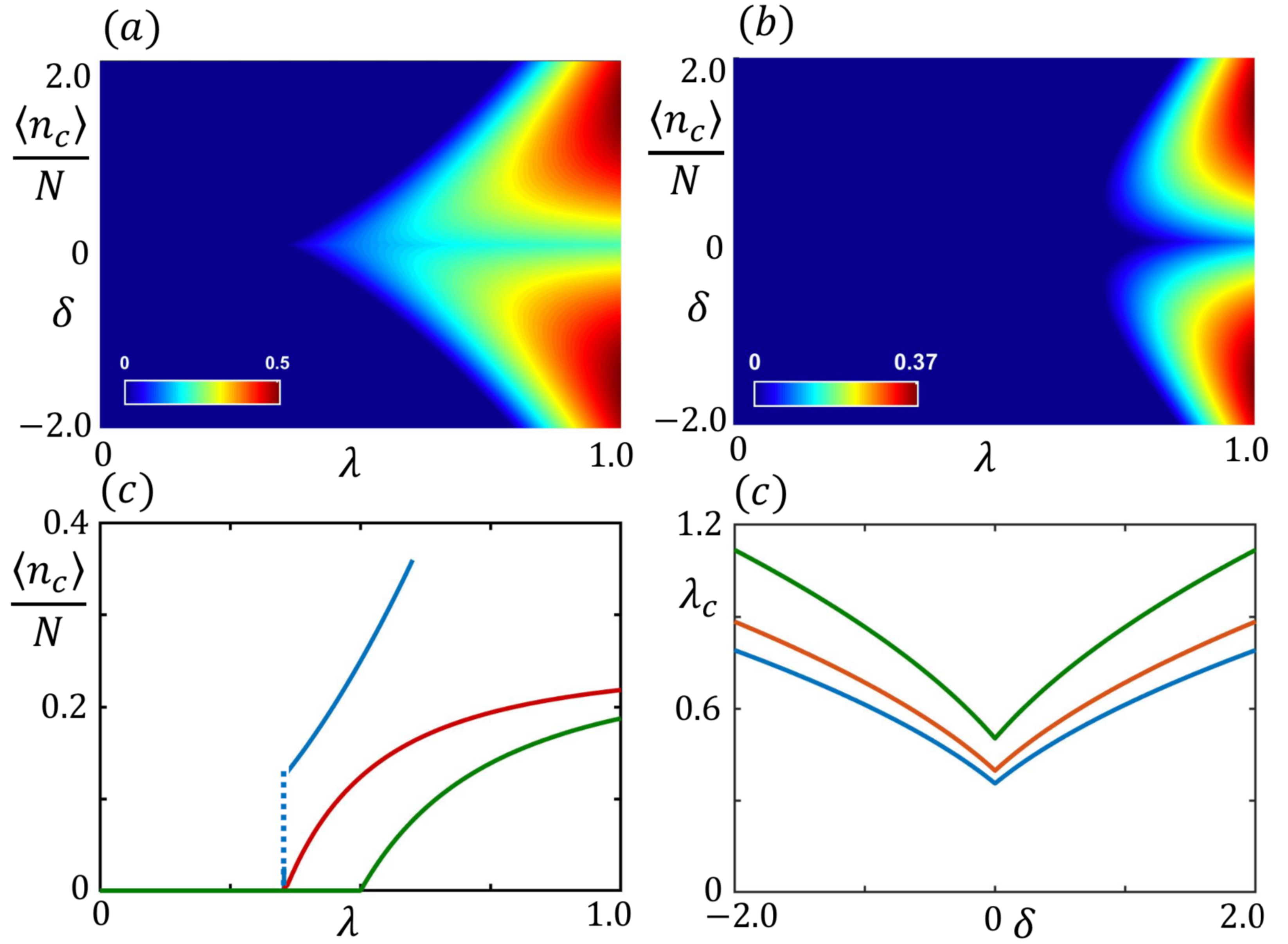}}
\caption{The expectation value of cavity photon number $\langle n_c\rangle/N$ for different $\delta$ and $\lambda$ in the system with (a). small atom decay rate $\gamma=0.1$ and (b). lager atom decay rate $\gamma=2$. (c). The expectation value of photon number $\alpha^2/N$ for different atomic emission rate $\gamma$.  $\gamma=0$ for blue line, $\gamma=0.1$ for red line, $\gamma=1$ for green line. (d). Phase boundry $\lambda_c$ for different parameters. $\kappa=0.1$ and $\gamma=0$ for blue line, $\kappa=0.5$ and $\gamma=0.1$ for red soild line, $\kappa=1$ and $\gamma=0.1$ for green soild line. All parameters are scaled to $\omega=q_r=m=1.0$. }
\end{center}
\end{figure}

The expectation of cavity photon number is shown in Fig. 5. There are some features of the phase diagram in the dissipative system. First, the loss of cavity photon has no qualitative influence to the phase diagram as shown in Fig. 5(d), because we just consider the case where $\omega\gg\kappa$. Second, the atom decay completely changes the properties around the line $\delta=0$. The first order phase transition along $\delta=0$ can not be observed any more for an arbitrary nonzero $\gamma$, as shown in Fig. 5(c). The expectation value of cavity photon number has also been suppressed around $\delta=0$. These new phenomena can be simply explained by the similar results in standard Dicke model. In the cavity-assisted SOC system, $\delta\approx 0$ means the effective energy detuning between two hyperfine spin states is very small, then the atom decay will dominate the physical properties of in this case. To observe the first order phase transition in this system, one should either reduce the spontaneous emission rate as far as possible or use two ground states with the same spontaneous emission rate from the excited state.

\section{The influence of nonlinear terms}

In real experiment, $Ua^{\dag}aS_z$ in equation (\ref{socham}) which coming from mutual feedback between cavity and atoms is always very small but not zero. If the influence of this nonlinear coupling is inavoidable, we should consider the following effective Hamiltonian
\begin{equation}
\begin{split}
\mathcal{H}_\mathrm{eff}=&\frac{p^2}{2m}N+\Big(\frac{pq_r}{m}+\delta\Big)S_z+\omega a^\dag a \\
&+\frac{\lambda}{\sqrt{N}}(a^\dag+a)(S_++S_-)+\frac{U}{N}a^\dag aS_z
\label{effhamsocmanybody2}
\end{split}
\end{equation} 
Comparing with the original Hamiltonian (\ref{effhamsocmanybody}), the COM momentum $p$ is still a good quantum number because the external trap is absent here. Although the inversion symmetry in Hamiltonian (\ref{effhamsocmanybody}) has been broken because of the nonlinear coupling, the more general inversion symmetry with translation $p\rightarrow -p$, $\delta\rightarrow -\delta$, $S_z\rightarrow -S_z$ and $U\rightarrow -U$ is still conserved, so we just need to consider the case of $U>0$. The analytical solution is hard to approach because of the nonlinear term, then we use the mean-field approximation to estimate the scale energy of Hamiltonian (\ref{effhamsocmanybody2}). With Holstein-Primakoff transformation $S_z=b^\dag b-N/2$, $S_+=b^\dag\sqrt{N-b^\dag b}$ and $S_-=\sqrt{N-b^\dag b}b$ where $b^\dag(b)$ is the creation (annihilation) operator with boson commutation relation $[b, b^\dag]=1$, the Hamiltonian (\ref{effhamsocmanybody2}) can be expressed as a Two-mode bosonic Hamiltonian. Then using two shifting boson operators $a^\dag=a_s^\dag+\sqrt{N}\alpha$, $b^\dag=b_s^\dag-\sqrt{N}\beta$, then using relation $\langle a_s\rangle=\langle b_s\rangle=0$, ignore the quantum fluctuation, the scale energy of Hamiltonian (\ref{effhamsocmanybody2})  $\epsilon(p,\alpha,\beta)=E(p,\alpha,\beta)/N$ can be estimated by
\begin{equation}
\begin{split}
\epsilon(p,\alpha,\beta)=&\frac{p^2}{2m}+\omega\alpha^2+\Big(\delta+\frac{pq_r}{m}\Big)\Big(\beta^2-\frac{1}{2}\Big)\\&-4\lambda\alpha\beta\sqrt{1-\beta^2}+U\alpha^2\Big(\beta^2-\frac{1}{2}\Big)
\label{sochamu}
\end{split}
\end{equation}
The minimum of the scale energy can be estimated numerically with the same method in above section. One of typical dispersion with $\delta=0$ and $\lambda_c=\sqrt{\omega E_r}/2$ is shown in Fig. 6(a). If $U=0$, there is a flat spectrum which inducing a series of novel critical phenomena. But in the case of $U>0$, the energy spectrum becomes asymmetric because of the inversion symmetry breaking. Then the energy minimum is at some position with $0<p_\mathrm{min}<q_r/2$. The mean-field phase diagram with $U>0$ is shown in Fig. 6(b) and two important influences of nonlinear term can be observed. First, there is no first order phase transition anymore because of the disappearance of the flat spectrum for arbitrary parameters. The expectation of cavity photons and magnetic quantum number of atom system changes continuously with both parameters. Second, the phase boundary is asymmetric, as the red line shown in Fig. 6(b). This asymmetry can be briefly explained as follow. Consider two points on phase diagram with $\delta=\pm\delta_0$ ($\delta_0>0$) and $\lambda=\lambda_c$. Without the influence of $U$, the phase boundary has the COM momentum $p_\mathrm{min}=\pm q_r/2$ with same ground state energy respectively because of the inversion symmetry. If $U>0$, because the effective energy detuning $\omega_0=\delta+pq_r/m$ has opposite signs at $p=\pm q_r/2$, the influence of $Ua^\dag aS_z$ has opposite contributions to the mean-field energy of above two cases as $U\alpha^2\langle S_z\rangle$, which increase (decrease) the ground state energy of superradiant state for $\langle S_z\rangle<0$ $(>0)$ respectively. So on the right side of phase diagram with $\delta=\delta_0>0$, the nonlinear term will enhance the superradiant phenomena and decrease the phase boundary, but vice versa on the left side with $\delta=-\delta_0<0$.

\begin{figure}[tph]
\begin{center}
{\includegraphics[width=0.47 \textwidth]{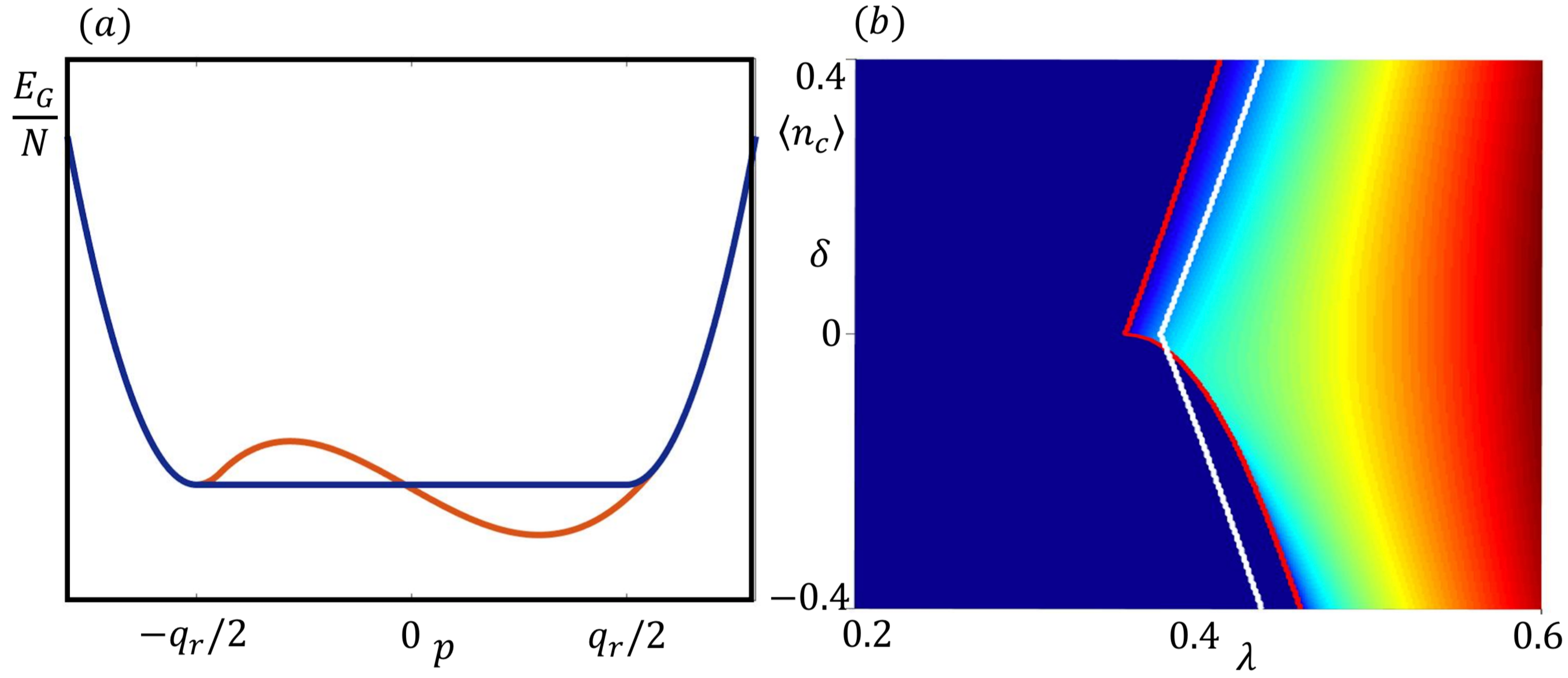}}
\caption{Phase diagram with the influence of nonlinear coupling $U$. (a). The energy spectrum with different COM momentum for $\delta=0$. the blue line corresponds to $U=0$ and the red line corresponds to $U>0$; (b). The expectation value of cavity photon number $\langle n_c\rangle$ when $U>0$, the red line represents the phase boundary, the white line represents phase boundary with $U=0$.  All parameters are scaled to $\omega=q_r=m=1.0$.}

\end{center}
\end{figure}

To further explain the influence of the nonlinear term, we can estimate the energy spectrum analytically using perturbation theory when $U$ is smaller than any other energy scales. In superradiant state, the expectation of excited atoms $\beta^2$ has the following form\cite{NLiu}
\begin{equation}
\beta^2=\frac{1}{2}-\frac{\omega}{U}+\frac{\omega}{U}\sqrt{\frac{4\lambda^2}{4\lambda^2+U\omega_0}\Big(1-\frac{U^2}{4\omega^2}\Big)}
\end{equation}  
where $\omega_0=\delta+\frac{pq_r}{m}$ which is dependent with the COM momentum $p$. Taking this expression into (\ref{sochamu}) and expanding the scale energy with $U$, we can get dispersion as
\begin{equation}
\begin{split}
\epsilon(p)\approx&p^2\Big(\frac{8m\lambda^2-\omega q_r^2}{16\lambda^2m^2}\Big)-\frac{\lambda^2}{\omega}+\Big(\frac{\omega^2\omega_0^2}{16g^4}-1\Big)\frac{U\omega_0}{8\omega}\\
\label{sochampu}
\end{split}
\end{equation}
which is a cubic function of COM momentum and asymmetric. When $\delta=0$ and $\lambda_0=\sqrt{\omega E_r}/2$, the dispersion will becomes $\epsilon(p)=-\frac{\lambda^2}{\omega}+\Big(\frac{\omega^2q_r^2p^2}{16g^4m^2}-1\Big)\frac{Uq_rp}{8\omega m}$ which is shown as the red line in Fig. 6(a).

\section{Discussion and Conclusion}

In conclusion, we discussed the influence of cavity induced dynamical spin-orbit coupling to ultracold atom systems. The phase diagram in thermodynamic limit shows some interesting performance. When two hyperfine spin states are degenerate, the energy spectrum at the phase boundary are flat and the excitation spectrum has the singularity. But from our discussion, this singularity is hard to be observed experimentally because of the atom decay in two-photon Raman process. We also discussed the influence of the nonlinear coupling which breaking the inversion symmetry makes the phase diagram asymmetric. 

The dynamical SOC effects induced by superrandiance opens a new way to quantum simulation in nonequilibrium systems. It has a wide application in the exploring the novel topological states in dynamical systems \cite{jspan,wzheng,mivehvar,colella}; the combination of the cavity-assisted SOC and the atomic interaction induces many interesting quantum phenomena in dynamic many-body systems \cite{kollath1,kollath2,ballantine,mivehvar2}; the dissipation of the cavity photons can also be utilized to explore the properties of open quantum systems\cite{dwwang,Gelhausen,yczhang,ostermann}. All these applications can be discussed in the future works and we hope the methods and results in this work can give some help to the following researches in this area.

\section*{ACKNOWLEDGEMENTS}
We thanks Han Pu and Xiang-Fa Zhou for their valuable suggestions. This research is supported by start-up funding of Huazhong University of Science and Technology and the Program of State Key Laboratory of Quantum Optics and Quantum Optics Devices (No: KF201903).

\begin{appendix}


\end{appendix}

\end{document}